\documentclass[reprint,amsmath,amssymb,aps]{revtex4-2}

\usepackage{graphicx}
\usepackage{bm}
\usepackage{hyperref}
\usepackage{amsmath}
\usepackage{braket}
\usepackage{dsfont}
\usepackage{siunitx}

\newcommand{\ee}{\mathrm{e}}
\newcommand{\ii}{\mathrm{i}}
\newcommand{\diff}{\mathrm{d}}

\begin{document}
	
    \title{Formation of the solid-state high-order harmonic generation plateau through destructive interference}

	\author{Lina Bielke}
	\author{Hannah J\"ur\ss}
	\author{Vincent Burgtorf}
	\author{Dieter Bauer}
	\affiliation{Institute of Physics, University of Rostock, 18051 Rostock, Germany }
	\date{\today}
	
	\begin{abstract}
        In frequently studied two-band models for solid-state high-harmonic generation, interband harmonics in principle can range from the minimum to the maximum bandgap. However, it is known that a laser-intensity dependent cutoff exists that may be well below the maximum bandgap unless the laser intensity is so high that the electrons explore the entire Brillouin zone. 	We show that this laser-intensity dependent cutoff is formed by destructive interference of the emission of electrons starting at different initial states in the Brillouin zone. The calculations in this work are for Su-Schrieffer-Heeger chains but our findings apply to other two-band systems as well. Only when the sampling of the Brillouin zone is fine enough or, equivalently, a finite chain is long enough in position space, the destructive interference is complete and forms the cutoff. For coarser sampling and shorter chains all harmonics between minimum and maximum bandgap are emitted.  A time-frequency analysis shows how certain trajectories are responsible for the formation of the cutoff.
	\end{abstract}

	\maketitle

	\section{Introduction} 
	    
	    High-order harmonic generation (HHG) is an important tool to create ultrashort laser pulses up to frequencies in the X-ray regime. It was first observed in atomic systems and subsequently described  using a semi-classical three-step-model \cite{corkum_plasma_1993,LewensteinPhysRevA.49.2117} 30 years ago. The efficient generation of high harmonics in solids was reported in 2010 \cite{Ghimire2011}. HHG in solids is very attractive for several reasons. For instance, the higher target density compared to gases allows for more compact radiation sources \cite{Luu2015,Ndabashimiye2016,LangerF.2017,HHG_Doped_MgO-Cr} and better scalability. From the fundamental research perspective, HHG can be used to probe static properties \cite{VampaPhysRevLett.115.193603,TancPhysRevLett.118.087403,Lakhotia2020} and dynamical processes \cite{SchubertO.2014,Hohenleutner2015,You2017,Baudisch2018,Tao-Yuan_2021_2, Uzan-Narovlansky2022} in condensed matter. HHG in solids has been recently reviewed in \cite{yue2021introduction,Goulielmakis2022}.
	    
	    In solids, several processes contribute to the generation of high-order harmonics: harmonics originating from the movement of electrons within a band (intraband harmonics) and the harmonics due to transitions between two bands (interband harmonics). Further, the time-dependent injection of electrons into the conduction band also generates harmonics \cite{Jurgens2020}. The well established three-step-model for atomic HHG can be adopted to describe the interband harmonics \cite{vampa_merge_2017}. First, an electron from the valence band is excited into the conduction band. This tunneling process occurs preferentially around the minimal band gap. In the presence of the laser-field, the electron (hole) moves in the conduction  (valence) band. Electron and hole might recombine, which leads to the emission of harmonic radiation.
	    
	    HHG in condensed matter can be theoretically simulated by solving the time-dependent Schr\"odinger equation (TDSE) for non-interacting electrons directly in position space (e.g., if finite-size effects and the influence of edges are of interest) or, after a Bloch ansatz, in momentum space for the bulk or a finite system with periodic boundary conditions. Electron-electron interaction can be included on a density-functional level \cite{TancPhysRevLett.118.087403}. The semiconductor Bloch equations can be used if relaxation processes due to couplings to other degrees of freedom (e.g., phonons or an environment) need to be taken into account via \textit{ad hoc} relaxation times \cite{vampa_merge_2017,haug_koch_quantum_2009}. Relaxation or dephasing processes were also modelled with an imaginary potential \cite{Tao-Yuan_2021}.
	    
	     In this work,  we restrict ourselves to the non-interacting TDSE level and a simple tight-binding description of a 1D chain, the so-called Su-Schrieffer-Heeger (SSH) model \cite{SSHPhysRevLett.42.1698} in order to investigate in detail a fact that we observed during our previous works: HHG spectra for the bulk might strongly depend on the $\vec k$-sampling of the Brillouin zone (BZ). For finite systems with periodic boundary conditions, the  $\vec k$-sampling is uniquely defined by the number of lattice sites $N$. However, for the bulk in the thermodynamic limit $N\to \infty$ one should in principle integrate over the first BZ or, numerically, sample fine enough to reach convergence. As we show in this paper, this convergence can be surprisingly slow, and a new qualitative feature in the HHG plateau, i.e., a pronounced drop in the harmonic yield,  emerges only for large enough $N$. This drop depends on the laser-intensity and can be seen as a cutoff. This is in agreement to other studies, e.g., \cite{Semiclassical_many_elec,gaarde_HHG,PhysRevX.7.021017,NavarretePRA2019} that report a similar effect. The same behavior is observed in TDSE simulations of large but finite chains in position space, showing that the observed effect is not merely a numerical curiosity but of physical relevance. 
	     
\begin{figure}
            \includegraphics[width = \linewidth]{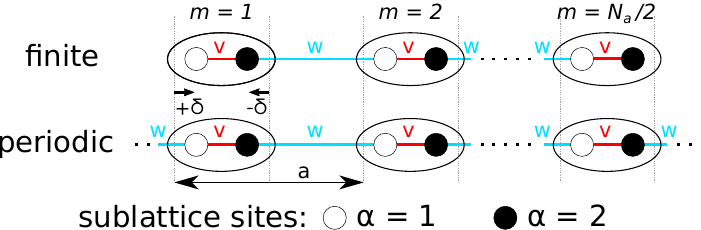}
            \caption{\label{fig1} Sketch of the SSH-model for a finite and the periodic chain. Each unit cell consists of two lattice sites, indicated by the open (sublattice site $\alpha = 1$) and filled (sublattice site $\alpha = 2$) circles. Unit cells are numerated by $m = 1,2,..., N_a/2$, with the total number of atoms $N_a$ in the chain chosen even. The hopping amplitude between sites within one unit cell is given by $v$ and across unit cells by $w$. The lattice constant is $a$. Relative to an equidistant grid (vertical lines), the lattice sites are shifted alternatingly by $\pm \delta$. For the periodic system, hopping between the right most site and the left most site of the chain is possible with amplitude $w$.}
        \end{figure}

	    
	     Figure \ref{fig1} shows a finite SSH-chain (top) and a periodic one (bottom).  The chains consists of two sublattice sites, indicated by open (sublattice site $\alpha = 1$) and filled (sublattice site $\alpha = 2$) circles. Two sites form a unit cell. The distance between two neighboring unit cells is the lattice constant  $a$. As the SSH-model is based on the tight-binding description, hopping amplitudes between adjacent sites are introduced. The hopping amplitude between sites in the same unit cell is $v$, and across neighboring unit cells it is $w$. Hopping (i.e., tunneling) should be more likely if two sites are closer together. Hence, different hopping amplitudes $v$ and $w$ correspond to different distances. Starting from the equidistant configuration where $v=w$ (dashed vertical lines), we consider the sites to be shifted alternatingly by $-\delta$ or $+\delta$. For the periodic system, hopping between the left most site and the right most site is allowed with amplitude $w$. Finite SSH chains  host topological edge states for $|w|>|v|$ \cite{asboth_book} whose effect on HHG has been studied in \cite{bauer_high-harmonic_2018, DrueekeBauer2019,JuerssBauer2019, Ma2022_HHG_SSH}. However, the topological nature of the SSH model is not relevant for the size and sampling dependence discussed in this work.
	
	    This paper is structured as follows. First, the methods and theory are described in Sec.  \ref{sec:methods}. The dependence of the HHG spectra on the size of the finite chains and the sampling of the BZ are shown in Sec.  \ref{sec:finite} and  \ref{sec:bulk}, respectively. Section \ref{sec:amplitude} covers the amplitude dependence of HHG before the results of a semi-analytical treatment for small laser field strengths are presented in Sec.  \ref{sec:analytical_small_A0}. In Sec.  \ref{sec:t-f-analysis}, a time-frequency analysis of the HHG is presented. The work is summarized in  Sec.  \ref{sec:summary}.
	
	\section{Methods}\label{sec:methods}
	    
	    Considering that tunneling probabilities scale exponentially with distance, we model the hopping amplitudes for the SSH chain as 
	    \begin{equation}
	        \begin{split}
	            v &= - \ee^{-(a/2 - 2\delta)}\\
	            w &= - \ee^{-(a/2 + 2\delta)}.
	        \end{split}
	    \end{equation}
	    The results shown in this paper are for a lattice constant of $a = 4.0$ and a shift of $\delta = 0.15$. However, our findings are  not specific  to this choice of model parameters ($a$ and $\delta$). 
	    
	    In the following, we briefly review the theory of the SSH bulk model and its coupling to a laser field. The treatment of finite SSH-chains is covered in \cite{JuerssBauer2019} and briefly summarized in the Appendix \ref{app:finite}.
	 
	    Atomic units ($\hbar = |e| = m_e = 4\pi\epsilon_0 =1$) are used unless stated otherwise.
	    
	    \subsection{Static bulk system}
	    
	        The bulk Hamiltonian of the SSH-chain can be obtained by a Bloch ansatz (see appendix \ref{app:Bloch} or \cite{Moos2020}), resulting  in a $2\times 2$-Hamiltonian
	        \begin{equation}\label{eq:H_0}
	            \hat{H}(k) = \begin{pmatrix}
	            	0 & s^*(k) \\
		            s(k) & 0 \\
	            \end{pmatrix}
	        \end{equation}
	        with
	        \begin{equation}
	            s(k) = v \ee^{-\ii (a/2 - 2 \delta) k} + w \ee^{\ii (a/2 + 2 \delta) k}.
	        \end{equation}
	            
	        Note that in Ref. \cite{Moos2020} the lattice constant $a$ is set to one and the shifts $\delta$ are considered small and hence set to zero. In this work we take the exact distances into account.
	        
	        The time-independent Schrödinger equation (TISE) for the bulk-system reads
	        \begin{equation}\label{eq:TISE}
            	E_n(k)
            	\mathbf{g}_n(k)
            	=
            	\hat{H}(k)
            	\mathbf{g}_n(k),
            \end{equation}	       
            where $\mathbf{g}_n(k) = \left(g_n^1 (k), g_n^2 (k)\right)^\top$, with $g_n^\alpha$ the value of the wavefunction at sublattice site $\alpha = 1,2$,  and the index $n = \pm$ indicates the valence ('$-$') or the conduction ('$+$') band. The band gap (energy difference between valence and conduction band for a given $k$) is given by 
            \begin{equation}\label{eq:band_gap}
                \begin{split}
                    E_g(k) &= E_+(k) - E_-(k) = 2E_+(k) = 2\sqrt{s(k)s^*(k)}\\
                    &=2\sqrt{v^2 + w^2 + 2vw\,\mathrm{cos}(ak)}.
                \end{split}
            \end{equation}

	    \subsection{Presence of an external field}
	    
	        In the presence of an external field, using velocity gauge and dipole approximation, the Hamiltonian becomes time-dependent
	        \begin{equation}\label{eq:H_bulk_t}
	            \hat{H}(k,t) = \begin{pmatrix}
                    0 & s^*(k,t) \\
                    s(k,t) & 0 \\
                \end{pmatrix}
	        \end{equation}
	        with 
	        \begin{align}
	            \begin{split}
	                s(k,t) &= v\ee^{-\ii (a/2 - 2 \delta) \left[k + A(t)\right]} + w \ee^{\ii (a/2 + 2 \delta) \left[k + A(t)\right]}\\ 
	                &= s(k+A(t)).
	            \end{split}
            \end{align}
	        Hence, the time-dependent function $s(k,t)$ is the time-independent function $s(k_\mathrm{eff})$ evaluated at an effective lattice momentum $k_\mathrm{eff} = k + A(t)$. As a consequence, the time-dependent Hamiltonian can be obtained by replacing the argument of the time-independent one, i.e.,  $\hat{H}(k,t) = \hat{H}(k + A(t))$. 
	    
	        A laser  pulse consisting of $n_\mathrm{cyc} = 5$ cycles is considered, described by the vector potential 
	        \begin{align}
	            A(t) = A_0\sin^2\left(\frac{\omega_0 t}{2 n_\mathrm{cyc}}\right)\sin\omega_0 t,~~~0<t<2\pi n_\mathrm{cyc}/\omega_0
            \end{align}	
	        and zero otherwise. The angular frequency  is set to $\omega_0 = 0.0075$ (i.e. the wavelength  $\lambda_0 \simeq 6.1~\mu\mathrm{m}$). The amplitude $A_0$ is varied in this paper but chosen positive. The laser pulse is quite short ($n_\mathrm{cyc} = 5$), hence one could think that the carrier-envelope-phase (CEP) might have an effect on our results. However, we have checked that the formation of the cutoff due to destructive interference is independent of the CEP.
 	    
	        The time-dependent Schrödinger equation reads
	        \begin{equation} \label{eq:ssh-bulk-bloch-matrix-time}
                \ii \mathbf{\dot{g}}(k,t) = \hat{H}(k,t) \mathbf{g}(k,t),
            \end{equation}
	        where $\mathbf{g}(k,t) = \left(g^1(k,t), g^2(k,t)\right)^\top$ is the time-dependent state in $k$-space.   The initial value is $\mathbf{g}(k,t=0) = \left(g_-^1(k), g_-^2(k)\right)^\top$, which is the eigenstate with the smallest energy for one particular $k$ (i.e., corresponding to the valence band) of the time-independent Hamiltonian (\ref{eq:H_0}).
	    
	        To obtain the harmonic spectrum, the current is calculated according to 
	        \begin{align}\label{eq:current_deriv}
                j(k,t) = \bm{g}^\dagger(k,t)\left[\partial_k \hat{H}(k,t)\right]\bm{g}(k,t).
            \end{align}
	        Hence, the derivative of the time-dependent Hamiltonian with respect to $k$ is the current operator. This statement only holds true if the right Bloch-ansatz is chosen in which the distances between the sites are taken into account \cite{Moos2020}. The derivative gives
	        \begin{align}
	            \partial_k \hat{H}(k,t) = \ii \begin{pmatrix}
                    0 & -s_-^*(k,t) \\
                    s_-(k,t)	& 0 \\
                \end{pmatrix}
	        \end{align}
	        with 
	        \begin{equation}
	            \begin{split}
                    s_-(k,t) =& (a/2 - 2 \delta) v\ee^{-\ii (a/2 - 2 \delta) \left[k + A(t)\right]}\\
                    - &(a/2 + 2 \delta) w \ee^{\ii (a/2 + 2 \delta) \left[k + A(t)\right]}.
	            \end{split}
            \end{equation}
	        The current (\ref{eq:current_deriv}) is calculated for different $k$-values within the first BZ with
	        \begin{equation}
	            k_n = n \frac{2 \pi}{N a}~~~\mathrm{and}~~~n \in [0,1,2,...,N-1].
            \end{equation}
	        Here, $N$ determines the sampling of the first BZ. The current is summed up over all calculated $k$-values, giving the total current
	        \begin{equation}\label{eq:total_current}
	            J(t) = \sum_{n=0}^{N-1}j(k_n,t). 
	        \end{equation}
	        We are interested in the spectrum of this current, which we obtain via Fourier transformation,
	        \begin{align}\label{eq:coherent}
	            I_\mathrm{total}(\omega) = \left|\int^{+\infty}_{-\infty} J(t)\, \ee^{-\ii \omega t}\mathrm{d}t\right|^2.
	        \end{align}
	        
	        In order to reveal interference effects it is useful to compare $I_\mathrm{total}(\omega)$ with the analogue incoherent result
	        \begin{equation}
	            I_\mathrm{incoherent}(\omega) = \sum_{n=0}^{N-1}I(\omega,k_n)
	        \end{equation}
	        in which the emission spectra of the single-electron currents with lattice momentum $k_n$ 
	        \begin{align}\label{eq:incoherent}
	            I(\omega,k_n) = \left| \int^{+\infty}_{-\infty} j(k_n,t)\, \ee^{-\ii \omega t}\mathrm{d}t\right|^2
	        \end{align}
	        are added up.

	        Fourier transforms are approximated by the fast Fourier transformation using a Hann-window and subtracting a constant offset to avoid artificial features.

	    \subsection{Time-frequency analysis}
	    
	        A time-frequency analysis displays the time-resolved emission spectrum. For this, the current  \eqref{eq:total_current} is first multiplied with a narrow window function $f(t,t_0)$, which is almost zero anywhere except in a close vicinity around $t = t_0$,
            \begin{equation}
                J(t,t_0) = J(t)f(t,t_0).
            \end{equation}
            Fourier-transforming this current gives $\tilde{J}(\omega,t_0)$ and the absolute value squared $|\tilde{J}(\omega,t_0)|^2$ reveals which harmonics are emitted at time $t_0$.
            The window function is chosen Gaussian, 
            \begin{equation}
                f(t,t_0) = \ee^{-(t-t_0)^2/(2\sigma)}
            \end{equation}
            with a full width at half maximum of FWHM $= 100$, which corresponds to $2\sigma \simeq 3606.7$.
            
            The variable $t_0$ is sampled in small steps between $t_0 = 0$ and the end of the laser pulse ($t_0 = 2\pi n_\mathrm{cyc}/\omega_0$). 
            For all Fourier transformations, a Hann-window as broad as the laser pulse is applied. 
            Whenever we show a  time-frequency analysis we replace $t_0$ by $ t$ at the abscissa.

	\section{Results}\label{sec:conv} 
	   
	\subsection{Convergence of HHG from finite SSH-chains}\label{sec:finite}
	
	    \begin{figure}
            \includegraphics[width = \linewidth]{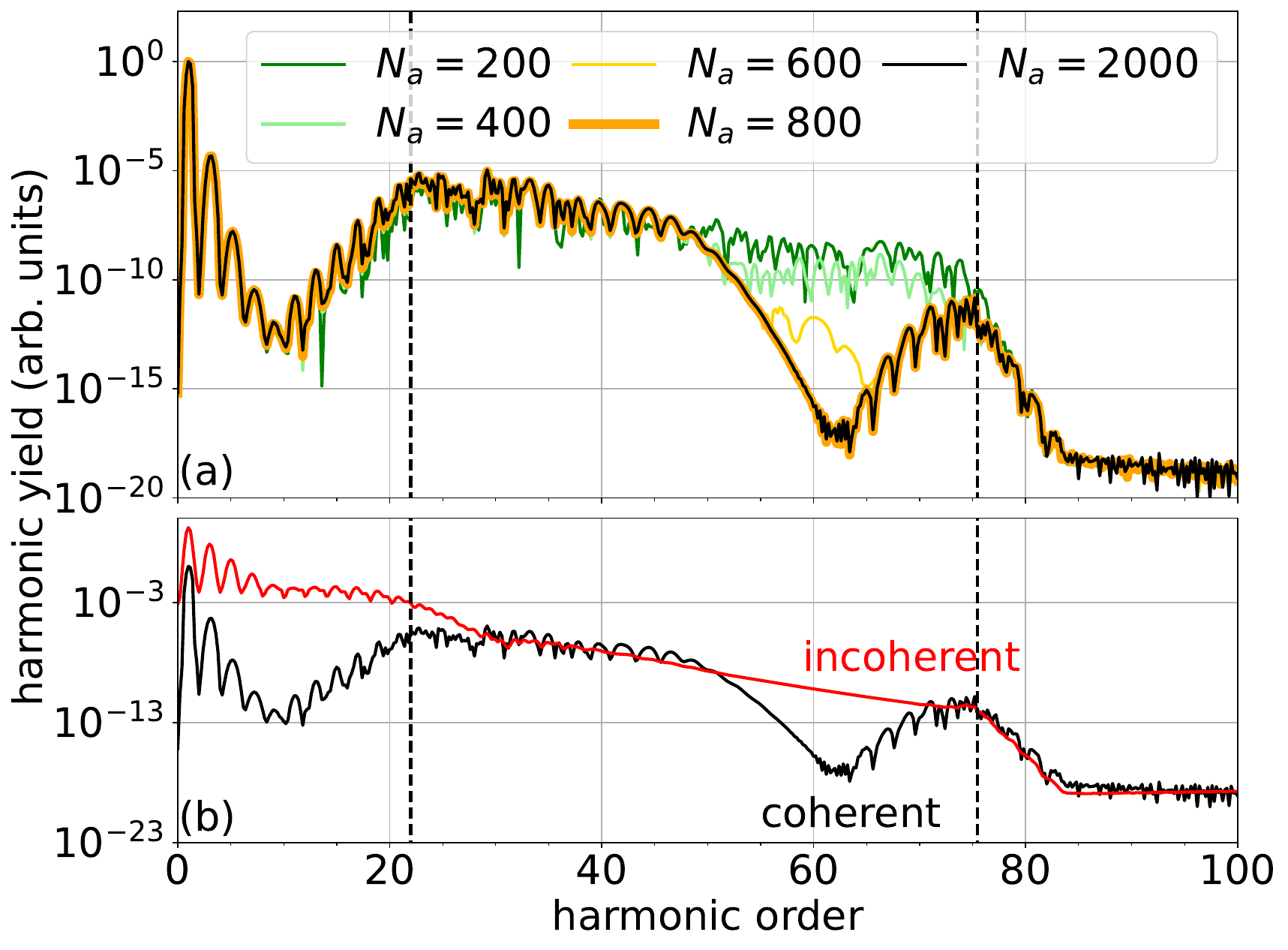}
            \caption{\label{fig2} (a) Harmonic yield for $A_0 = 0.2$ and different finite chain sizes. Spectra are normalized to their maximum. (b) Harmonic yield for a coherently and incoherently summed up dipole acceleration for $A_0 =0.2$ and $N_a = 2000$. The dashed vertical lines in (a) and (b) indicate the minimal and maximal band gap.}
        \end{figure}
	
	    High-harmonic radiation emitted from finite chains depend on the size of the chains, see for instance Ref. \cite{PhysRevA.97.043424}. In previous studies, harmonic spectra of a finite chain containing $100$ sites were investigated using time-dependent density functional theory \cite{bauer_high-harmonic_2018,DrueekeBauer2019} and the SSH-model \cite{JuerssBauer2019}.  The results were similar for chains containing $30$ and $100$ sites \cite{DrueekeBauer2019}. 
	
	    For our current work, we calculated HHG spectra for much longer, finite chains. In Fig.\  \ref{fig2}a, the spectra for different numbers of atoms $N_a$ are shown. For the spectra of the finite system the dipole acceleration (see appendix \ref{app:finite}) is used instead of the current (used for the bulk results). We do this to increase the dynamic range such that the high-frequency emission is not buried under the background due to the finite-time Fourier transform. The amplitude of the vector potential is $A_0 = 0.2$ (corresponding to a laser intensity of $\simeq 7.9 \cdot 10^{10}$ W/cm$^{2}$). The spectra for $N_a = 200$ and $N_a = 400$ are similar. However, if $N_a$ is increased further, differences appear. This is seen for $N_a = 600$ and even more obvious for $N_a = 800$ where a drop in the harmonic yield around order $60$ clearly developed. Changing the number of atoms to a much higher number of $N_a = 2000$ does not change the spectrum any further.
	    The point is that if we only considered chains up to  $N_a = 200$ and $N_a = 400$ we might had erroneously concluded that convergence had been reached already. Note that the effect we observe here is very different from the finite-size effects discussed in \cite{PhysRevA.97.043424}, in which much smaller chains (up to $100$ sites) are investigated. 
	    
	    The drop in the harmonic yield causes a cutoff at a smaller energy for longer chains compared to smaller ones. The increased yield around the maximal band gap is already several orders smaller than the harmonic yield of the plateau. It is not negligible but far less important than the plateau. 
	
	    In Fig. \ref{fig2}b,  the harmonic yield for the coherently and incoherently summed up dipole acceleration for $N_a = 2000$ sites are compared. Only the result for the coherent sum shows the drop in the harmonic yield around harmonic $60$. As the incoherent sum does not show this feature, it has to originate from destructive interference of the emissions due to single-electron currents.
	
	\subsection{Convergence of HHG from the bulk of the SSH-chain}\label{sec:bulk}

        \begin{figure}
            \includegraphics[width = \linewidth]{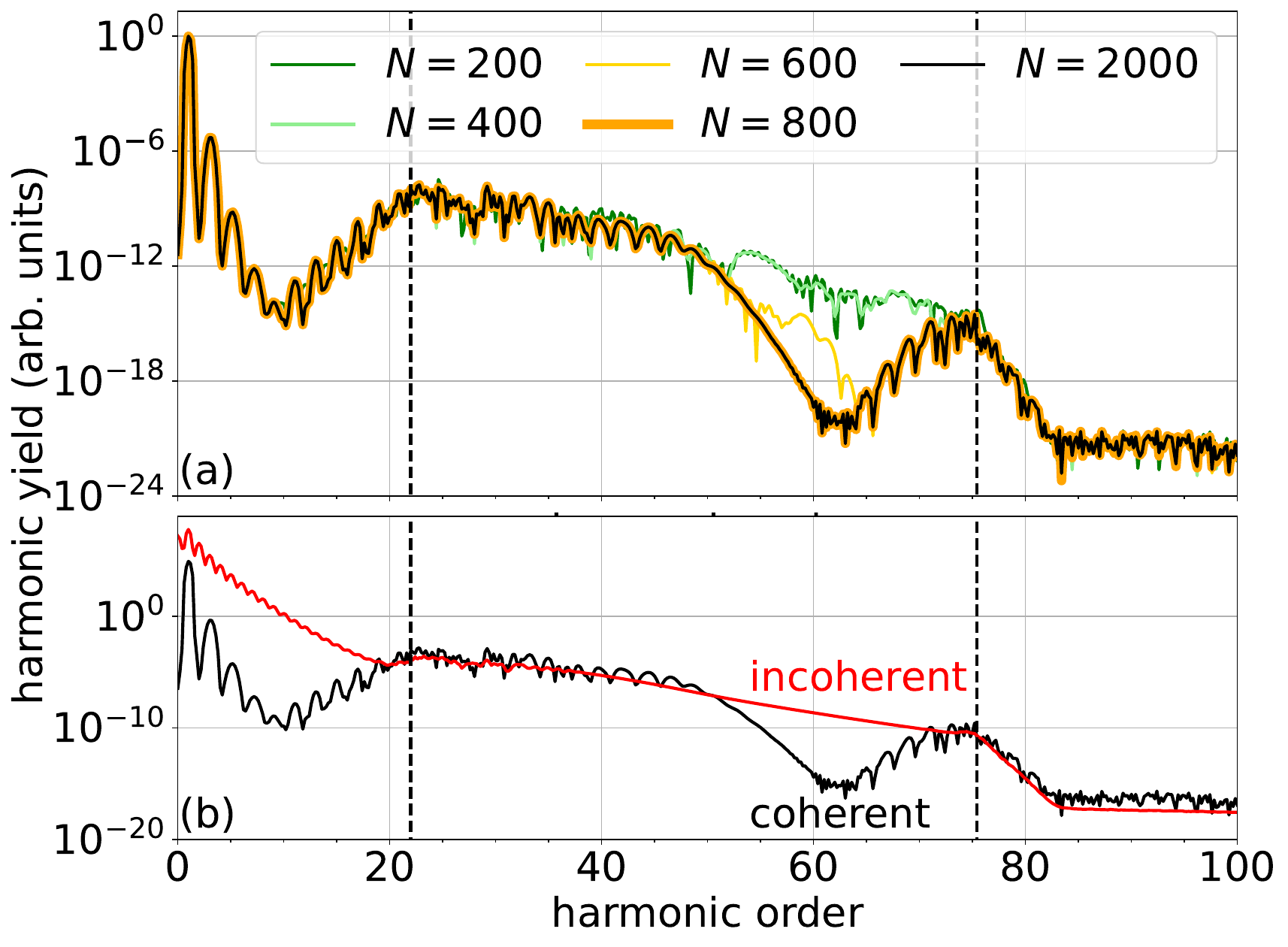}
            \caption{\label{fig3} (a) Harmonic spectra ($A_0 = 0.2$) for different sampling rates of the first BZ of the bulk, normalized to their maximum values.  (b) Harmonic yield for a coherently and incoherently summed up current for $A_0 = 0.2$ and $N = 2000$. The dashed vertical lines in (a) and (b) indicate the minimal and maximal band gap.}
        \end{figure}
        
        HHG spectra in the bulk limit are calculated according to  \eqref{eq:coherent} with the total current \eqref{eq:total_current}.
        In Fig. \ref{fig3}a, the spectra for different samplings $k_n$, $n=0,1,2,\ldots N-1$ are shown for a laser amplitude of $A_0 = 0.2$. The spectra are similar to the results for the finite chain: if the BZ is not sampled fine enough (small $N$), the spectrum has a plateau for interband harmonics between the minimal and maximal band gap (dashed lines). If the sampling is higher, the harmonic yield around harmonic order $60$ decreases significantly. This leads effectively to a smaller plateau with a cutoff around harmonic order $50$. A local maximum around the maximal band gap is observed as well.
        This decreased harmonic yield before the maximal energy gap is again due to destructive interference, as it is not visible for the calculation using the incoherent sum, see Fig. \ref{fig3}b. In the bulk, the different single-electron currents originate from different $k$-points within the first BZ. The emissions from those interfere destructively. 
        
        The study of the dependence of the HHG spectra on the sampling appears to be a convergence test and, as such, rather technical. However, a discrete sampling of the BZ corresponds to a finite system in position space with periodic boundary conditions. Further, in the previous section we investigated, explicitly in position space, finite chains with different sizes. This is not a convergence test, as real physical systems are finite and can have different sizes. The observed spectroscopic dip in the interband harmonics appears in both finite-size position space and BZ bulk results. It is thus not only a numerical effect. In the following we investigate converged spectra for the bulk system. The bulk system has the advantage that its band structure is known analytically.

    \subsection{Dependence on the amplitude}\label{sec:amplitude}

	       \begin{figure}
                \includegraphics[width = \linewidth]{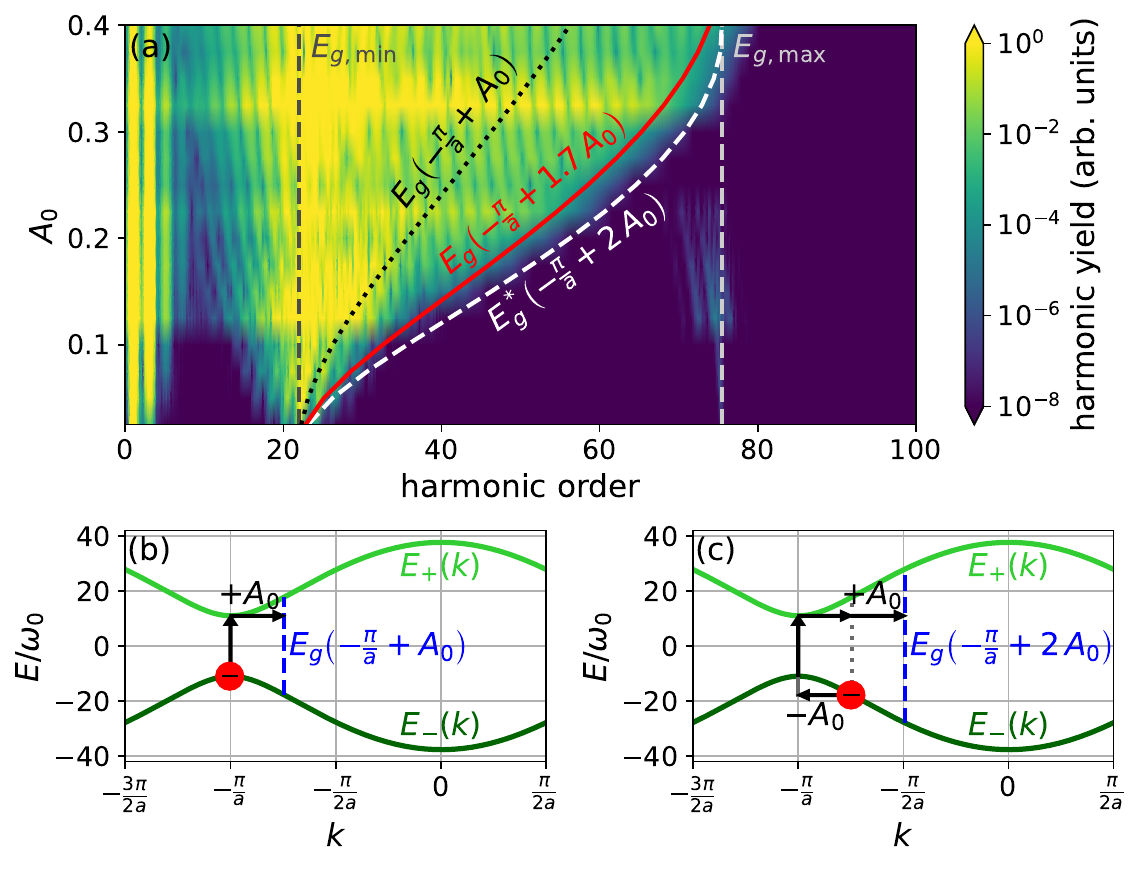}
                \caption{\label{fig4} (a) High-harmonic spectrum as function of $A_0$ for $N = 1000$ normalized to the yield at the minimal band gap $E_{g,\mathrm{min}}$. The vertical lines indicate the minimal ($E_{g,\mathrm{min}}$) and maximal ($E_{g,\mathrm{max}}$) band gap. The black dotted line $E_g(-\frac{\pi}{a} +A_0)$ indicates the maximal band gap that can be reached if an electron tunnels at the minimal band gap when $A(t) = 0$ (shown in (b)). The white dashed line $E^*_g(-\frac{\pi}{a} +2 A_0)$ indicates the maximal band gap that can be reached if the electron tunnels at the minimal band gap when the vector potential is minimal (shown in (c)). The red line $E_g(-\frac{\pi}{a} + 1.7 A_0)$ corresponds to a value in between, the tunneling occurs when $A(t) = -0.3 A_0$. (b, c) Sketch of the process within the band structure leading to interband harmonics. }
            \end{figure}
            
            Figure \ref{fig4}a shows high-harmonic spectra as a function of the amplitude of the vector potential $A_0$ for the bulk and $N = 1000$ sampling points in the first BZ. The spectra are normalized to the intensity of the harmonic corresponding to the minimal band gap. A plateau can be observed for energies larger than the minimal band gap. The cutoff of the plateau depends on the laser intensity and shifts to larger harmonic orders as the intensity increases. A dependence of the cutoff on the amplitude of the laser was expected, as shown for example in Refs. \cite{Semiclassical_many_elec,gaarde_HHG,PhysRevX.7.021017,NavarretePRA2019}. Around the maximal band gap ($\simeq 75 ~\omega_0$), a local maximum is observed. However, this yield is much smaller than the yield in the region of the plateau (see also Fig. \ref{fig3}) and hence not visible in this contour plot for $0.25 <A_0 < 0.3$.
            
            For a given $k$, the probability for electrons to be excited from the valence into the conduction band depends on the energy gap and the respective transition matrix element. Excitation is typically more likely if the energy gap is small. Therefore, most electrons tunnel into the conduction band at the minimal band gap \cite{vampa_merge_2017}. In the presence of the laser field, the electrons move inside the bands according to $k(t) = k_0 + A(t)$, where $k_0$ is the $k$-value for $A(t) = 0$. 
	        
	        If the electron tunnels into the conduction band when the vector potential is zero, the electron can move at most $A_0$ to the left or to the right, see Fig. \ref{fig4}b. The minimal band gap is located at $k = -\frac{\pi}{a}$. As it was shown in Ref.\ \cite{Gaarde_imperfect_recol}, the electron and hole do not necessarily have to recollide in position space in order to recombine and emit light. Assuming that the electron might recombine with the hole in the valence band at any time, the highest possible transition energy is  $E_g(-\frac{\pi}{a} + A_0)$. Here, $E_g(k)$ is the band gap at momentum $k$, see  (\ref{eq:band_gap}). If the electron is displaced too far from the minimal band gap, it might reach the maximal band gap (at either $k = -2\,\frac{\pi}{a}$ or $k = 0$). Hence,  if $A_0 > \frac{\pi}{a}$ the electron-hole pair will be shifted beyond the maximal band gap, and the maximal transition energy is given by the maximal band gap. The black dotted line in Fig. \ref{fig4}a shows the function $E_g(-\frac{\pi}{a} + A_0)$. This line does not agree well with the cutoff frequency. 
	        
	        Alternatively, the electron might also tunnel at a different phase of the vector potential. The most extreme case would be when the absolute value of the vector potential is maximum. Figure \ref{fig4}c shows the case for tunneling at the minimum of the vector potential. In this case, the electron is initially located at a $k$-value given by $k_0 = -\frac{\pi}{a} + A_0$. This electron is driven to the minimal band gap (at $k = -\frac{\pi}{a}$) when $A(t) = -A_0$. After the electron has tunneled, it is driven back to its original position in $k$-space when the vector potential is zero. At the maximum vector potential, the electron-hole pair is $2\, A_0$ away from the minimal band gap. Hence, the highest transition energy for this scenario is given by $E_g(-\frac{\pi}{a} + 2 A_0)$ (unless $2\, A_0 > \frac{\pi}{a}$, in which case the maximal band gap is the largest transition energy possible). The dashed white line in Fig. \ref{fig4}a indicates the value of the function
            \begin{equation}
                E^*_g(-\frac{\pi}{a} + 2 A_0) = \begin{cases}
                E_g(-\frac{\pi}{a} + 2 A_0)&\text{if $2\, A_0 < \frac{\pi}{a}$}\\
                E_g(0)&\text{else}
                \end{cases}
            \end{equation}
	        where $E_g(0)$ is the maximal band gap $E_{g,\mathrm{max}}$.
	        This function agrees better with the cutoff, although not precisely as the harmonic yield at this energy is already several orders below the yield in the plateau. The given examples for certain tunneling instants are only the extreme cases. The electron might also tunnel at any other phase of the vector potential. However, the maximal displacement of the electron from the minimal band gap is given by $c\, A_0$ with the parameter $c$ restricted to values $1 \leq c \leq 2$. The red solid line in Fig. \ref{fig4}a indicates the energy difference for $c = 1.7$ ($E_g(-\pi/a +1.7\,A_0)$), which appears to be a good fit to describe the cutoff as function of $A_0$. Previously found cutoff laws show a linear dependence for various materials \cite{Semiclassical_many_elec,gaarde_HHG,PhysRevX.7.021017,NavarretePRA2019}. This linear scaling is different from ours because of the almost linear dispersion relation in the relevant part of the BZ for these materials.
	        
	        We also found that the position of the cutoff is not dependent on the number of laser-cycles  $n_\mathrm{cyc}$, which was tested for $A_0 = 0.2$ up to $n_\mathrm{cyc} = 20$ laser-cycles (not shown).

	   \subsection{Semi-analytical approximation for low intensities}\label{sec:analytical_small_A0}
        
            \begin{figure}
                \includegraphics[width = \linewidth]{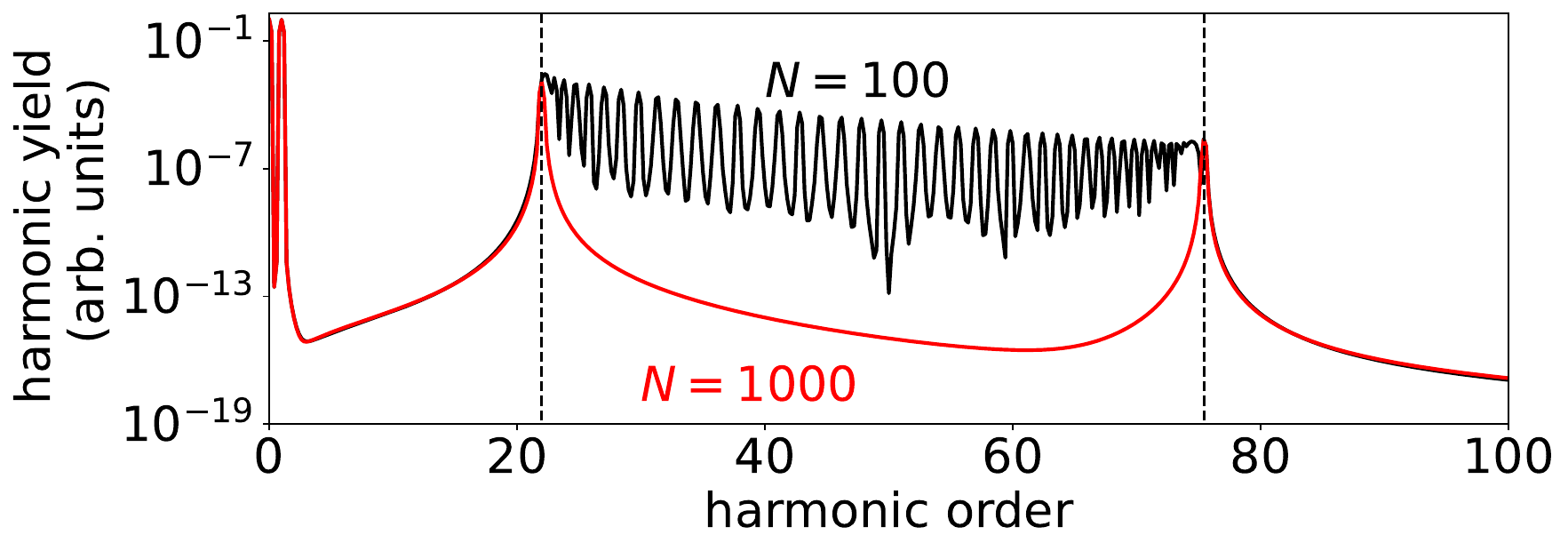}
                \caption{\label{fig5} Solution of the semi-analytical approximation of the total current for weak fields (here: $A_0 = 0.01$) normalized to their maximum harmonic yield. The dashed vertical lines indicate the minimal and maximal band gap energy.}
            \end{figure}
        
            We have seen already that the cutoff energy of the harmonic spectrum depends on the vector potential and is due to destructive interference of the emission due to single-electron currents from different $k$-values. In this section, the occurrence of the destructive interference is investigated mathematically. To that end, a small laser intensity is considered.
        
            The analytical expression for the interband current in the bulk SSH-model is given in \cite{Moos2020}. This expression is still not solvable analytically. However, if we consider a weak laser field ($A_0 \simeq 0$) we can approximate $k(t) = k + A(t) \simeq k$.  In Appendix \ref{app:analytical_calc}, the calculation for this approximation is performed. The final result is 
            \begin{align}\label{eq:analytic_smallA}
                j(k,t) &\propto  \frac{(w^2-v^2)^2}{4E^3_+(k)} \left\{2\,A_0\cos(\omega_0 t) -  A_0 \, \cos[E_g(k)t]\right\}.
            \end{align}
            The total current is given by integration over the first BZ
            \begin{align}\label{eq:J}
                J(t) = \int_{\mathrm{BZ}}j(k,t)\,\diff k.
            \end{align}
            This integral is solved numerically, using  $N$ sampling points within the first BZ.
            Figure \ref{fig5} shows the results for different $N$. For a small number of sampling points ($N = 100$), a peak at the fundamental harmonic and a plateau at higher energies can be observed. For a larger number of points ($N = 1000$), \eqref{eq:J} gives only three peaks, at the fundamental harmonic and both the minimal and maximal band gap. The current in  (\ref{eq:analytic_smallA}) obviously generates the fundamental harmonic $\omega_0$ due to the term with $\mathrm{cos}(\omega_0 t)$ and harmonics between minimal and maximal band gap because of the term  $\mathrm{cos}(E_g(k)t)$. In the integrated current,  however, only the fundamental harmonic and peaks around the minimal and maximal band gap survive. This result agrees with the simulation presented in Fig. \ref{fig4}a when the amplitude of the vector potential $A_0$ is small. 
            
            For larger vector potentials, this approximation is not applicable anymore and the equations become too complicated if higher-order terms in the Taylor expansion of $E_+(k(t))$ are included. As the result in Fig. \ref{fig4}a shows, the peak around the minimal band gap expands towards higher energies. The peak at the maximal band gap expands slightly towards smaller energies but this is hardly  visible. Harmonics between those energies still interfere destructively. The semi-analytical calculation clarifies the origin of the decreased harmonic yield between the minimal and maximal band gap but does not predict the development of the spectrum with increasing laser intensity.

	\subsection{Time-frequency analysis}\label{sec:t-f-analysis}
	   
       \begin{figure}
            \includegraphics[width = \linewidth]{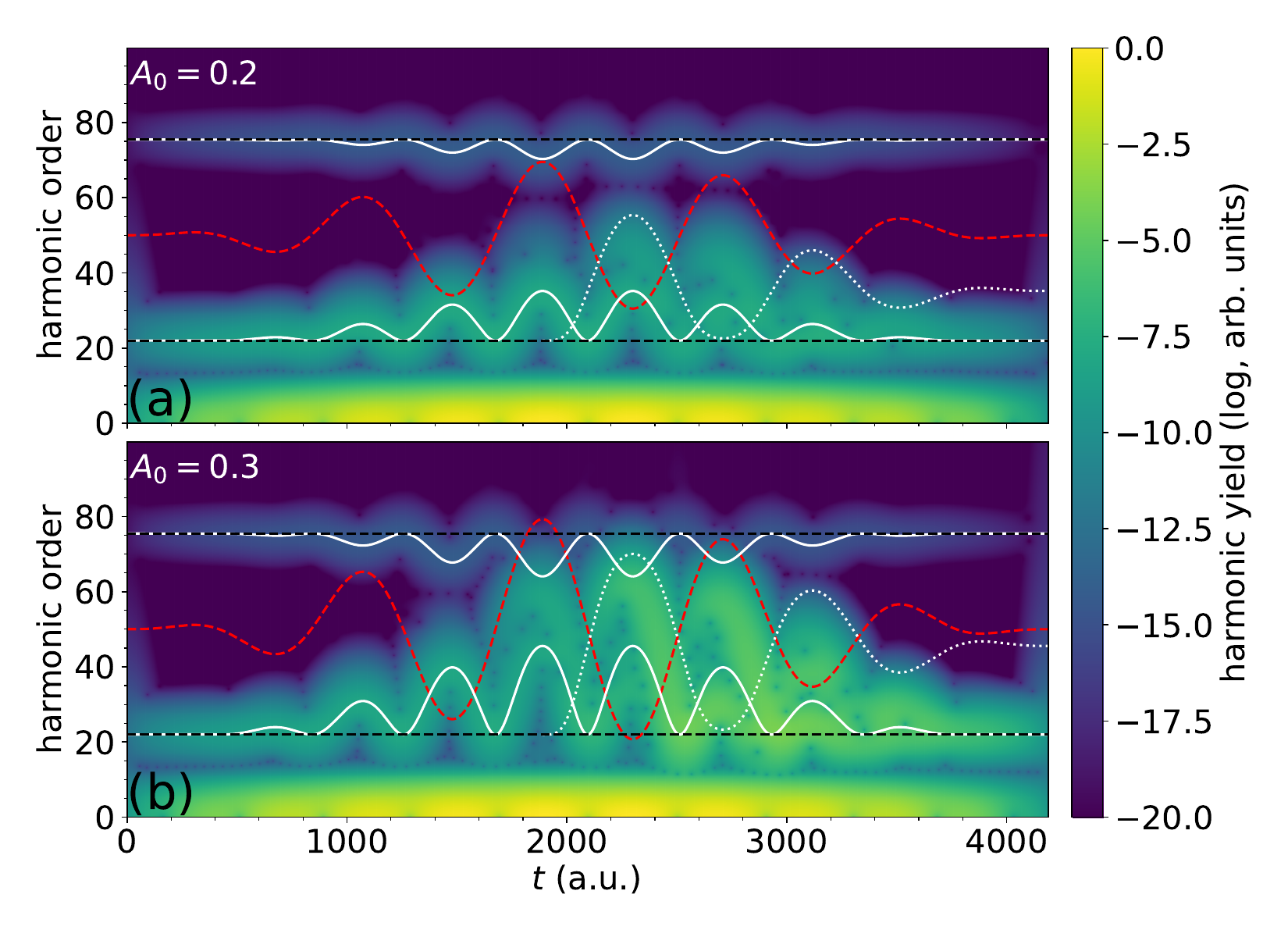}
            \caption{\label{fig6} Time-frequency analysis for different laser intensities: $A_0 = 0.2$ (a) and $A_0 = 0.3$ (b) for a sampling of $N = 1000$. The colorbar is in logarithmic scale (with basis $10$), each plot is normalized to its maximum value. The red dashed line indicates the shape of the vector potential (units on $y$-axis suppressed). The horizontal dashed lines indicate the maximal and minimal band gap energy. The white lines (solid and dotted) indicate different trajectories.}
        \end{figure}

        A time-frequency analysis is performed to gain insight into the dynamical processes underlying the harmonic spectra and finally the cutoff law that was found empirically in Sec.   \ref{sec:amplitude}.

	   Figure \ref{fig6} shows the result of the time-frequency analysis for different field amplitudes for a sampling rate of $N = 1000$ (converged results). The spectra shown are the integrated spectra over the whole BZ. Arc-like structures are visible for energies between the minimal and maximal band gap (black dashed, horizontal lines).  
	 
	   The clear signatures in the time-frequency analysis that correlate certain harmonics with certain emission times indicate that there must exist an explanation in terms of semi-classical orbits \cite{Lein_physRevA_2010, PhysRevLett.105.203902,Semiclassical_many_elec, PhysRevA.93.023402,PhysRevA.100.043404, yue2021introduction, Tao-Yuan_2021_2}. In the following, we try to identify those semi-classical orbits.

	   The dashed, red line indicates the shape of the vector potential.  As discussed in Sec.   \ref{sec:amplitude}, the electrons move inside the conduction band with the vector potential $k(t) = k_0 + A(t)$. Assuming vertical transitions, the electrons in the conduction band can recombine with the holes in the valence band and emit a photon with the respective energy difference between both bands at the given $k(t)$. The white lines indicate the transition energy $E_g(k(t))$ at the respective $k(t)$ over time. The two solid white lines are for electrons that start at either the maximum or minimum band gap at the beginning of the laser pulse. The dotted white line indicates the case where the electron tunnels into the conduction band at the minimal band gap when the vector potential is close to its maximum.   
	   
	   Around the maximal band gap, an oscillating pattern is observed, best seen for the smaller amplitude in Fig. \ref{fig6}a. The white solid line agrees well with these oscillations. This line indicates the case where   the tunneling process happens at the maximal band gap when the vector potential $A(t)$ is zero. This is expected, as tunneling processes are more likely when the electric field $-\partial_t A(t)$ is large, i.e., $A(t) \simeq 0$. 
	   
	   Around the minimal band gap, more and stronger oscillations and arches are observed. The solid white line close to the minimal band gap in Figs. \ref{fig6}a, b shows the transition energies over time if the electron tunnels into the conduction band at the minimal band gap when the vector potential $A(t)$ is zero. Parts of the emission spectrum can be explained by this pathway but not all. The dotted white line indicates the transition energy when the tunneling process occurs close to the maximum of the vector potential (but still at the minimal band gap). These electrons can explore more of the band structure, which leads to higher possible transition energies, as discussed in Sec. \ref{sec:amplitude} (see also Fig. \ref{fig4}c). Therefore, these pathways contribute to the higher harmonics. 
	   
	   Not all features around the minimal band gap  can be explained by those two trajectories alone.  The agreement between the semi-classical trajectories and the time-frequency analysis may  be improved by performing a more detailed calculation using the saddle-point approximation \cite{Lein_physRevA_2010, Semiclassical_many_elec}. 
	   
	   These findings suggest that for the converged results only electrons which tunnel into the conduction band around either the minimal or maximal band gap contribute to the emission spectrum. The harmonics in between cancel out due to destructive interference if the sampling of the BZ is fine enough. Around the maximal band gap, the tunneling process requires a high electric field, i.e.,  a  vector potential close to zero. Around the minimal band gap, however, tunneling is much more likely and can also occur for much smaller values of the electric field (larger $A(t)$). Hence, trajectories of electrons which tunnel when the vector potential is large also contribute to the overall spectrum and are responsible for high-energetic photons.

       Comparing the results for different laser intensities shows the development of the cutoff law. Only certain trajectories contribute to the overall spectrum, and other harmonics cancel out due to destructive interference. In particular, harmonics that can be explained by semiclassical trajectories do not cancel out if the electrons are excited into the conduction band around the minimal band gap. When the intensity is small, the electrons cannot move away from the minimal band gap much, leading to a cutoff at a rather small energy. When the intensities increase, the electrons can move further away from the minimal band gap. This leads to a larger possible transition energy in this semi-classical picture. Hence, the cutoff shifts to larger energies.

    \section{Summary and conclusion}\label{sec:summary}
    
        In this work, we found a laser-intensity dependent cutoff in the high-harmonic spectra from the SSH-chain. In addition, a local maximum around the maximal band gap is observed. This cutoff in the harmonic yield is only observed in the total harmonic spectrum taking into account the emission by all electrons. Instead, the cutoff is absent in single-electron or incoherently added harmonic spectra. Hence, the cutoff forms due to the destructive interference of single-electron emission from different $k$-values. This effect is observed in both finite and periodic SSH-chains. However, the chains have to be very long (or the sampling of the BZ very high) to see this effect. With a too coarse sampling of the BZ or too short finite chains, the cutoff is not obtained and the plateau harmonics range from the minimal to the maximal band gap without formation of a cutoff in between. 
        
        One may object that a tight-binding model with just two bands is too simplistic to be of practical relevance. However, a cutoff that depends on the laser-intensity was confirmed experimentally \cite{Ghimire2011} and also found in many other theoretical works, e.g.,  \cite{Semiclassical_many_elec,gaarde_HHG,PhysRevX.7.021017,NavarretePRA2019}.  Further, the simple modelling has also advantages. Note, for instance, that our results are strictly gauge invariant with respect to the coupling to the laser field \cite{Graf_1995} and the choice of a basis (e.g., field-free or adiabatically following). This is different from the case where the number of bands is restricted \textit{after} the Bloch ansatz for the (continuous) Schr\"odinger equation was made and a basis was chosen \cite{Yue2020}.  Only the somewhat counter-intuitive  local maximum in the harmonic yield around the maximum band gap might be due to the limitation to only two bands. 
        
        Another critical issue is that decoherence due to coupling to other degrees of freedom (e.g., phonons or environment) might spoil the delicate destructive-interference effect observed in this work. In fact, studies showed that the decreased harmonic yield for intraband harmonics, found for example in Refs. \cite{bauer_high-harmonic_2018,JuerssBauer2019}, is not observed or less prominent if a finite,  \textit{ad hoc} dephasing time is introduced \cite{Yue2020,Ma2022_HHG_SSH} in the Lindblad or semiconductor Bloch equations for the  density matrix. Whether the  decrease in the interband harmonic yield found in our present study survives dephasing will be examined in future work. Due to numerous other works that show an amplitude-dependent cutoff in HHG from solids (e.g. \cite{Semiclassical_many_elec,gaarde_HHG,PhysRevX.7.021017,NavarretePRA2019}), we assume the effect can still be observed to a certain extend if dephasing is included. Additionally, the before mentioned local maximum in the harmonic yield around the maximal band gap may be destroyed or weakened if dephasing is included. Anyhow, it is useful to first understand the relatively simple and ``clean'' time-dependent Schr\"odinger results before introducing relaxation times.
    
        Finally, we note that we also found a drop in the harmonic yield at certain energies for the two-dimensional Haldane model \cite{Haldane_1988} if the sampling of the BZ was sufficiently high. It thus appears that the effect is rather general and not restricted to the one-dimensional SSH-model.

	\section*{Acknowledgment}

	    Funding by the German Research Foundation - SFB 1477 “Light-Matter Interactions at Interfaces,” Project No. 441234705, is gratefully acknowledged. H.J. acknowledges financial support by the doctoral fellowship program of the University of Rostock.

    \begin{appendix}

    \section{Methods for the finite SSH-chain}\label{app:finite}
    
        This section briefly summarizes the methods used for calculating harmonic spectra for finite SSH-chains. The equations are similar to Ref. \cite{JuerssBauer2019}. However, note that in that reference $a$ denotes the distance between neighboring sites. In this paper, $a$ is the lattice constant.
        
        \subsection{Static system}
        
            The Hamiltonian of the unperturbed system (without external field) for an even number of atoms $N_a$ in the chain reads
            \begin{equation}\label{eq:H_finite} 
                \begin{split}
                \hat{H}_{\mathrm{finite}} &= \sum_{m=1}^{N_a/2} \left(v \ket{m,2} \bra{m,1}\right) \\
                &+ \sum_{m=1}^{N_a/2-1} \left(w \ket{m+1,1} \bra{m,2} \right) + \mathrm{h.c.}.
                \end{split}
			\end{equation}
            The TISE
            \begin{equation}
                 \hat{H}_{\mathrm{finite}} \Psi_i = E_n \Psi_i,
            \end{equation}
            is solved numerically. Here, the eigenstate $\Psi_i$ has the form $\Psi_i=\left(\Psi_i^{1,1},\Psi_i^{1,2},...,\Psi_i^{m,\alpha},...,\Psi_i^{N_a/2,1},\Psi_i^{N_a/2,2}\right)^\top$. The vector component $\Psi_i^{m,\alpha}$ is the value of the wavefunction of state $\Psi_i$ at sublattice site $\alpha$ in unit cell $m$ (see Fig. \ref{fig1}). There are $N_a$ eigenstates of the TISE (i.e., $i = 0,1,2,...,N_a-1$), which can be sorted according to their energies, $E_0 \leq E_1 \leq E_2 \leq ... \leq E_{N_a-1}$.  
            
            The absolute positions of the sites are chosen
            \begin{equation}\label{eq:positions}
                x_{m,\alpha} = m\,a + \left(a/2 - 2\delta \right)\delta_{\alpha,2}, \quad m = 1,2,...,N_a/2
			\end{equation}
			with the Kronecker delta $\delta_{\alpha,2}$. With this convention, the position of the first site (see Fig. \ref{fig1}) is set to $x_{m=1,\alpha = 1} = 0$.
			
		\subsection{Coupling to an external field}
		    
		    For the coupling to an external field, the velocity gauge is used. In that gauge the hopping elements become time-dependent (Peierls substitution) \cite{Graf_1995}
		    \begin{align}
		        v(t) &= v\exp[{-\ii(a/2-2\delta)A(t)}] \nonumber \\ 
		        &= -\exp\{-(a/2-2\delta)[1+\ii A(t)]\}, \\
                w(t) &= w \exp[{-\ii(a/2+2\delta)A(t)}] \nonumber \\ 
                & = -\exp\{-(a/2+2\delta)[1+\ii A(t)]\}. 
			\end{align}
			The time-dependent Hamiltonian has a similar structure as the time-independent one (\ref{eq:H_finite}) but with the time-dependent hopping amplitudes
			\begin{equation} \label{eq:H_finite_t}
                \begin{split}
                \hat{H}_{\mathrm{finite}}(t) &= \sum_{m=1}^{N_a/2} \left(v(t) \ket{m,2} \bra{m,1} + v^*(t)\ket{m,1}\bra{m,2} \right) \\
                &+ \sum_{m=1}^{N_a/2-1} \left(w(t) \ket{m+1,1} \bra{m,2} \right.\\
                &\left. + w^*(t) \ket{m,2}\bra{m+1,1} \right).
                \end{split}
			\end{equation}
			
			The TDSE
			\begin{equation}
			    \ii\partial_t \Psi_i(t) = \hat{H}_{\mathrm{finite}}(t) \Psi_i(t)
			\end{equation}
			is solved numerically using the Crank-Nicolson approximation with the initial value given by the eigenstates of the unperturbed Hamiltonian $\Psi_i(t = 0) = \Psi_i$. The TDSE is solved for all states $i = 0,1,2,..., N_a/2-1$ corresponding to the valence band. The other states correspond to the initially unpopulated conduction band.
			
			Instead of the current, the dipole acceleration is used to calculate the harmonic spectrum for finite chains. References \cite{PhysRevA.79.023403,0953-4075-44-11-115601} showed that the harmonic spectrum can be calculated using either the dipole, the current, or the dipole acceleration. For sufficiently long pulses, the results differ  by prefactors $\omega^ 2$.  The total dipole is given by
			\begin{equation} 
			    X(t) = \sum_{i = 0}^{N_a/2 -1}\sum_{m=1}^{N_a/2} \Psi_i^{m*}(t) x_m \Psi_i^m(t).
		    \end{equation}	
		    The harmonic spectra shown in Fig. \ref{fig2}a, b was calculated as the absolute value squared of the Fourier-transformed dipole acceleration $\ddot{X}(t)$. We chose the dipole acceleration here to increase the dynamic range compared to Fourier transforms of the current or the dipole itself. 
			
			\section{Bloch-ansatz for the bulk-Hamiltonian}\label{app:Bloch}
    
                Starting point for the derivation of the Hamiltonian of the bulk system (\ref{eq:H_0}) is the tight-binding Hamiltonian of the periodic system with $N_a$ atomic sites in position space
                \begin{equation}\label{eq:H_periodic} 
                    \hat{H}_{\mathrm{bulk}} = \sum_{m=1}^{N_a/2} \big( v \ket{m,2} \bra{m,1} + w \ket{m+1,1} \bra{m,2} \big) + \mathrm{h.c.}
			    \end{equation}
    			with the periodic boundary condition $\ket{N_a/2 + 1,\alpha} = \ket{1,\alpha}$.
    			In order to solve the TISE
    			\begin{equation}\label{eq:TISE_periodic}
                    \hat{H}_{\mathrm{bulk}} \ket{\Psi_n} = E_n \ket{\Psi_n},
                \end{equation}
                we make the Bloch-ansatz
                \begin{equation} \label{eq:ssh-bulk-ansatz}
                	\begin{split}
                		\ket{\Psi_n(k)} &= \frac{1}{\sqrt{N_a}} \sum_{m=1}^{N_a/2} \exp (\ii m a k) \ket{m}\\
                		&\times\sum_{\alpha=1,2}\exp \{\ii (a/2- 2\delta)k\delta_{\alpha,2} \} g_n^{\alpha} (k) \ket{\alpha},
                	\end{split}
                \end{equation}
                in which the positions of all sites are considered, see (\ref{eq:positions}). This is important to ensure physically meaningful currents when using  \eqref{eq:current_deriv}. 
            
                One obtains
                \begin{equation} \label{eq:ssh-bulk-eigenvalue1}
                	\begin{split}
                		\hat{H}&_{\mathrm{bulk}} \ket{\Psi_n(k)} = \frac{1}{\sqrt{N_a}} \sum_{m=1}^{N_a/2}
                		g_n^2\exp (\ii m a k) \\
                		&\times\exp\{\ii (a/2 - 2 \delta) k\}\left[ v \ket{m,1} + w  \ket{m+1,1} \right] \\
                		&+g_n^1\exp (\ii m a k) \left[ v \ket{m,2} + w \exp (\ii  a k) \ket{m,2} \right],
                	\end{split}
                \end{equation}
                and 
                \begin{equation} \label{eq:ssh-bulk-eigenvalue2}
                	\begin{split}
                		\sqrt{N_a}&\bra{m'}\hat{H}_{\mathrm{bulk}} \ket{\Psi_n(k)} =
                		g_n^2\exp (\ii m' a k) \\
                		&\times\exp\{\ii (a/2 - 2 \delta) k\}\left[ v + w \exp (-\ii  a k) \right]\ket{ 1}  \\
                		&+g_n^1\exp (\ii m' a k) \left[ v + w \exp (\ii  a k)\right]\ket{2}.
                	\end{split}
                \end{equation}
                Analogously, 
                \begin{equation} \label{eq:ssh-bulk-eigenvalue3}
                	\begin{split}
                		\sqrt{N_a}&\bra{m'}E_n(k) \ket{\Psi_n(k)} = E_n\exp (\ii m' a k)\\
                		&\times\left[g_n^{1} (k) \ket{1} + \exp \{\ii (a/2- 2\delta)k\} g_n^{2} (k) \ket{2}\right].
                	\end{split}
                \end{equation}
                Hence, from the TISE \eqref{eq:TISE_periodic} follows
                \begin{equation}
                	\begin{split}
                		\sqrt{N_a}\bra{m'}\hat{H}_{\mathrm{bulk}} \ket{\Psi_n(k)} = \sqrt{N_a}\bra{m'}E_n(k) \ket{\Psi_n(k)},
                	\end{split}
                \end{equation}
                and in matrix form 
                \begin{equation}
                	E_n(k)
                	\begin{pmatrix}
                		g_n^1 (k) \\
                		g_n^2 (k) \\
                	\end{pmatrix}
                	=
                	\begin{pmatrix}
                		0		& s^*(k) \\
                		s(k)	& 0 \\
                	\end{pmatrix}
                	\begin{pmatrix}
                		g_n^1 (k) \\
                		g_n^2 (k) \\
                	\end{pmatrix},
                \end{equation}
                with
                \begin{equation}
                	s(k) = v \exp(-\ii (a/2 - 2 \delta) k) + w \exp(\ii (a/2 + 2 \delta) k).
                \end{equation}
                This gives the Hamiltonian for the bulk (\ref{eq:H_0}). A similar calculation can be performed to derive the time-dependent Hamiltonian in (\ref{eq:H_bulk_t}).
			
			\section{Semi-analytical solution for small vector potentials}\label{app:analytical_calc}
			
			    An analytical approximation for the interband harmonics is given by the real part of the equation (66) in Ref. \cite{Moos2020}
            	\begin{align}
            	       \begin{split}
            	           v^{-+}(k,t) &\simeq   \frac{(w^2-v^2)^2}{8E_+(k+A(t))}\int^t  \diff t' \, \frac{\dot A(t')}{E_+^2(k+A(t'))}  \\ &\times\ee^{-\ii \int^t_{t'}2E_+(k+A(t'')) \,\diff t'' }
            	       \end{split}
                \end{align}
            	where $E_+$ is the dispersion relation for the conduction band.
            	
            	For a weak field  $A(t)$, the time-dependent momentum is approximated as $k(t) = k + A(t) \simeq k$, that is, it is time-independent. Further, the expression $E_+(k + A(t)) \rightarrow E_+(k)$ is now time-independent as well. As a consequence, 
            	\begin{align}
            	    \begin{split}
            	        v^{-+}(k,t) &\simeq \frac{(w^2-v^2)^2}{8E^3_+(k)}\int^t  \diff t' \, \dot A(t') \ee^{-2\ii E_+(k)\int^t_{t'} 1 \,\diff t'' }\\
            	        &= \frac{(w^2-v^2)^2}{8E^3_+(k)}\int^t  \diff t' \, \dot A(t')  \ee^{-2\ii E_+(k)\,(t-t')},
            	    \end{split}
            	\end{align}
                which can be solved with integration by parts,
                \begin{align}
                    \begin{split}
                    &\int^t  \diff t' \, \dot A(t') \,\ee^{-2\ii E_+(k)\,(t-t')} = \left[A(t')\,\ee^{-2\ii E_+(k)\,(t-t')} \right]_{t' = 0}^t \\
                    &- \int^t  \diff t' \, A(t') \,\ee^{-2\ii E_+(k)\,(t-t')}2\ii E_+(k).
                    \end{split}
                \end{align}
                The first term on the right-hand side gives $A(t)$ (because we assume that the vector potential is zero in the beginning $A(t' = 0) = 0$) such that it 
                only contributes to the fundamental harmonic $\omega_0$. 
            
                The interband current is proportional to 
                \begin{align}
                    \begin{split}
                        &j(k,t) \propto 2\Re(v^{-+}(k,t)) \simeq 2\Re\left[\frac{(w^2-v^2)^2}{8E^3_+(k)}\right.\\
                        &\left.\times \left(A(t) - \int^t  \diff t' \, A(t') \,\ee^{-2\ii E_+(k)\,(t-t')}2\ii E_+(k) \right)\right] \\
                        &= \frac{(w^2-v^2)^2}{4E^2_+(k)}\left(\frac{A(t)}{E_+(k)} \right.\\
                        &\left.-2\int^t  \diff t' \, A(t') \,\mathrm{sin}\left(2 E_+(k)\,(t-t')\right)\right) .
                    \end{split}
                \end{align}
                We assume that the intensity of the laser pulse is slowly ramped up so that the pulse envelope can be neglected. Considering  $A(t) = A_0\,\mathrm{cos}(\omega_0\,t)$, we obtain
                \begin{widetext}
                    \begin{align}
                    \begin{split}
                    j(k,t) &\propto  \frac{(w^2-v^2)^2}{4E^2_+(k)}\left(\frac{A(t)}{E_+(k)} - 2A_0\int^t  \diff t' \, \mathrm{cos}(\omega_0\,t) \,\mathrm{sin}\left(2 E_+(k)\,(t-t')\right)\right)\\ 
                    & = \frac{(w^2-v^2)^2}{4E^2_+(k)}\left(\frac{A(t)}{E_+(k)} + A_0 \left[ \frac{\mathrm{cos}((2E_+(k) - \omega_0)t' - 2E_+(k)t)}{2E_+(k) - \omega_0} + \frac{\mathrm{cos}((2E_+(k) + \omega_0)t' - 2E_+(k)t)}{2E_+(k) + \omega_0} \right]_{t' = 0}^t\right)\\
                    & = \frac{(w^2-v^2)^2}{4E^2_+(k)}\left(\frac{A(t)}{E_+(k)} + A_0  \left[ \frac{\mathrm{cos}(- \omega_0 t) - \mathrm{cos}(- 2E_+(k)t) }{2 E_+(k) - \omega_0} + \frac{\mathrm{cos}( \omega_0 t) - \mathrm{cos}(- 2E_+(k)t)}{2E_+(k) + \omega_0} \right]\right)\\
                    &\simeq \frac{(w^2-v^2)^2}{4E^3_+(k)} \left(A(t) +  A_0 \left[ \mathrm{cos}( \omega_0 t) - \mathrm{cos}(2E_+(k)t)  \right]\right) = \frac{(w^2-v^2)^2}{4E^3_+(k)} \left(2\,A(t) -  A_0 \, \mathrm{cos}(2E_+(k)t)\right).
                    \end{split}
                    \end{align}
                \end{widetext}
                In the last line, we assumed that the driver frequency is much smaller than the band gap $\omega_0 \ll 2E_+(k)=E_g(k)$, as is usually the case in HHG. The total current follows by integration over the first BZ
                \begin{align}
                    J(t) = \int_{\mathrm{BZ}}j(k,t)\,\diff k,
                \end{align}
                which we performed numerically.

    \end{appendix}

	\bibliography{biblio.bib}

\end{document}